\documentclass{iopart}
\usepackage{graphicx,amssymb,amsfonts,units,bm} 

\begin{document}

\title{Non-Commutative Tools for Topological Insulators}

\author{Emil Prodan}

\address{Department of Physics, Yeshiva University, New York, NY 10016}

\begin{abstract}
This paper reviews several analytic tools for the field of topological insulators, developed with the aid of non-commutative calculus and geometry. The set of tools includes bulk topological invariants defined directly in the thermodynamic limit and in the presence of disorder, whose robustness is shown to have non-trivial physical consequences for the bulk states. The set of tools also includes a general relation between the current of an observable and its edge index, relation that can be used to investigate the robustness of the edge states against disorder. The paper focuses on the motivations behind creating such tools and on how to use them.
\end{abstract}


\date{\today}

\maketitle

\section{Introduction}

The topological insulating state represents an entirely novel manifestation of the ordinary matter, characterized by the existence of metallic edge states despite the presence of strong disorder or strong deformations of the materials. We will argue here that the topological insulating state is also characterized by the existence of bulk states that resist localization. These are intrinsic properties of the topological insulators and no external magnetic field is need to trigger this peculiar behavior \cite{Murakami:2004dx}. 

Fervent theoretical activity around the topological insulators resulted in a variety of characterizations, some based on very original ideas  \cite{Kane:2005zw,Fu:2006ka,Murakami:2006bw,Sugimoto:2006zr,Fu:2007ti,Fu:2007vs,Teo:2008zm,Sheng:2006nq,Sheng:2006na,Moore:2007ew,Murakami:2007zd,Murakami:2007mo,Essin:2007ij,Qi:2008cg,Moore:2008ok,Roy:2009cc,Roy:2009am}. For example, the Z$_2$ invariant introduced by Kane and Mele \cite{Kane:2005zw} was new to the physics and mathematics communities. Same can be said about the Spin-Chern number introduced by Sheng et al \cite{Sheng:2006na} or about the recent results on 3D topological insulators. We also want to mention that several topological invariants appeared long before the field of topological insulators was defined, in the context of $^3$He-A like topologically ordered systems \cite{Volovik:1989eb}. 

Most of the analytic works were completed within the framework of translational invariant bulk systems and homogeneous edges, where one can build topological invariants on the Brillouin torus and the edge states can be computed explicitely. The topological properties in the presence of disorder were mostly investigated by computational means \cite{Sheng:2005mg,Chen:2005bx,Sheng:2005um,Sheng:2005lj,Sheng:2006nq,Sheng:2006na,Essin:2007ij,Onoda:2007xo}. The present contribution to this special issue on "Topological Insulators" gives a short overview of the author's recent works \cite{Prodan:2009od,Prodan:2009lo,Prodan:2009mi,Prodan:2009oh}, which provide a set of analytic tools to diagnose the effect of disorder on the bulk and on the edge states of the topological insulators. 

The topic of the present paper is how to define topological invariants directly in the thermodynamic limit and in the presence of disorder and what kind of information can be extracted from their invariant properties. The discussion excludes technical arguments, which can be found in the author's published work, but includes a careful presentation of the assumptions and the limits in which the results are valid. The connection to other works, especially to the existing numerical studies, is emphasized.  The paper also pays special attention to the motivation behind these works and how to use the emerging analytic tools.  The author also shares some of his current personal beliefs that crystalized during this work. 

\section{Tools for the bulk states}

We start the discussion from the classical Chern invariant. For any smooth family of projectors $\{P_{\bm \theta}\}_{{\bm \theta}\in {\bm M}}$ indexed by ${\bm \theta}$, who lives on a two dimensional smooth and closed manifold ${\bm M}$, one can compute the Berry curvature
\begin{equation}\label{Berry}
dF =(2\pi i)^{-1} \mbox{Tr} \{P_{\bm \theta}[\partial_{\theta_1}P_{\bm \theta},\partial_{\theta_2}P_{\bm \theta}]\}d\theta_1 \wedge d \theta_2
\end{equation}
and then integrate the curvature over ${\bm M}$ to obtain the Chern number, an integer that is invariant to smooth deformations of the family of projectors. We want to emphasize from the beginning that the Chern number can be defined only if $P_{{\bm \theta}}$ is at least first order differentiable in ${\bm \theta}$, for all ${\bm \theta}$ of the manifold ${\bm M}$. The Chern number can change its value if $P_{{\bm \theta}}$ seize to be a global smooth function of ${\bm \theta}$.

The last two observations are quite relevant because many invariants for topological insulators were linked to the Chern numbers of families of projectors $P_{\bm k}$, built through various procedures, without any check on the global smoothness of $P_{\bm k}$ with respect to ${\bm k}$ (the Bloch wave-vector). Without such check, the constructions remain formal and incomplete. The issue should not be treated lightly since proving global smoothness of various objects on the Brillouin torus represents an outstanding problem in the band theory. Physicists as well as mathematicians bumped into it when trying to extend Kohn's results on the analytic structure of Bloch functions from one to higher dimensions  \cite{Nenciu:1991qa,Prodan:2006pt,Brouder:2007p233,Panati:2007sy,Thonhauser:2006kx}. What usually happens is the following: the objects are easily seen to be locally smooth on ${\bm k}$ and, starting from a small region of the Brillouin torus, the objects can be smoothly extrapolated to larger and larger regions. But when it is to cover the entire torus, point singularities in the first derivatives can develop that cannot be removed by any gauge transformation. Proving the absence of such singularities is almost always a formidable task \cite{Prodan:2006pt}. 

The construction of the spin-Chern number given in Ref.~\cite{Prodan:2009oh} includes a check of the global smoothness of the projectors involved in the construction. Let us repeat the construction of Ref.~\cite{Prodan:2009oh}, this time emphasizing this check. We will use the concrete model of electrons in graphene \cite{Kane:2005np, Kane:2005zw}:
\begin{equation}\label{model}
\begin{array}{c}
H_0=-t\sum\limits_{\langle {\bm m \bm n} \rangle,\sigma} |{\bm m},\sigma\rangle \langle {\bm n},\sigma| \medskip \\
+i\lambda_{SO}\sum\limits_{\langle \langle {\bm m \bm n} \rangle \rangle,\sigma\sigma'}  [ {\bf s} \cdot ({\bf d}_{\bm k \bm m} \times {\bf d}_{\bm n \bm k} )]_{\sigma,\sigma'} |{\bm  m},\sigma\rangle \langle {\bm  n},\sigma'| \medskip \\
+i\lambda_R\sum\limits_{\langle {\bm m \bm n} \rangle,\sigma \sigma'}  [ \hat{{\bf z}}\cdot ({\bf s}\times {\bf d}_{\bm m \bm n})]_{\sigma,\sigma'} |{\bm n},\sigma\rangle \langle {\bm n},\sigma'|.
\end{array}
\end{equation}
Here, {\bf s} is the spin operator, ${\bm m}$ and ${\bm n}$ denote the sites of the honeycomb lattice and $\sigma$ and $\sigma'$ the electron spin degrees of freedom, taking the values $\pm 1$. The Hamiltonian acts on the Hilbert space ${\cal H}$ spanned by the orthonormal basis $|{\bm n},\sigma\rangle$. The simple angular brackets in Eq.~\ref{model} denote the nearest neighbors and the double angular brackets denote the second nearest neighbours. Inside the second sum, ${\bm k}$ represents the unique common nearest-neighbor of ${\bm m}$ and ${\bm n}$. Also, ${\bf d}_{\bm m \bm n}$ is the displacement from the site ${\bm m}$ to the site ${\bm n}$, with the units chosen so that the distance between nearest neighbors is unity.  The electrons are considered non-interacting. The three terms in Eq.~\ref{model} are the usual nearest neighbor hopping term, the intrinsic spin-orbit coupling preserving the lattice symmetries and the Rashba potential induced by the substrate supporting the graphene sheet. We assume that the parameters in the Hamiltonian are chosen so that we are in the Spin-Hall zone of the phase diagram. At some point in the paper, we will view the honeycomb lattice as a triangular lattice of unit cells containing two C atoms. When we do so, we will use the notation $|{\bm i},\alpha,\sigma\rangle$ for the elementary quantum states, where ${\bm i}$ represents the position of a unit cell in the triangular lattice and $\alpha=1,2$ is a label for the two atoms inside each unit cell.  

There are two sites per unit cell and two spin states per site, therefore the model has 4 states per unit cell. For the beginning, we will stay with the translational invariant case, where we can use the Bloch fibration, i.e., the unitary transformation from the Hilbert space of the infinite sample into a continuum direct sum of $4$ dimensional complex spaces:
\begin{equation}
\begin{array}{c}
U: {\cal H} \rightarrow \bigoplus_{{\bm k}\in {\cal T}} {\bm C}^4, \ \ (U{\bm \Psi})({\bm k}) = \frac{1}{2\pi} \sum_{\bm j} e^{-i {\bm k}\cdot{\bm j}} {\bm \Psi}({\bm j}),
\end{array}
\end{equation}
where ${\bm k}$ lives on the Brillouin torus ${\cal T}=[0,2\pi]$$\times$$[0,2\pi]$. $U$ transforms the original Hamiltonian in a direct sum of Bloch Hamiltonians: $UH_0U^{-1} = \bigoplus_{{\bm k}\in {\cal T}} H_0({\bm k})$, which display two  upper and two lower bands separated by an insulating gap. Let $P$ denote the spectral projector onto the states below the insulating gap of $H_0$. Under the Bloch fibration, this projector becomes: $UPU^{-1}$=$\bigoplus_{{\bm k}\in {\cal T}}P({\bm k})$ and the Chern number associated with the family of projectors $P({\bm k})$ is zero, as it will generically be for any time reversal invariant band model. According to Ref.~\cite{Panati:2007sy}, the fiber bundle of the occupied states is topologically equivalent to the trivial fiber bundle ${\cal T}\times {\bm C}^2$.

It is clear that, in order to obtain non-trivial Chern numbers, we need to split the projector $P({\bm k})$=$P_-({\bm k})$$\oplus$$P_+({\bm k})$ and hope that $P_\pm({\bm k})$ have non-trivial Chern numbers. Due to the time-reversal invariance and the half-integer value of the spin, one can always split $P({\bm k})$ like this and there are in fact an infinite way of doing the splitting at each ${\bm k}$. This observation was used, for example, in Refs.~\cite{Roy:2009cc} and \cite{Roy:2009am} to give alternative methods of calculus for the Z$_2$ invariant. But the big question is: Among this infinite ways of splitting, is there at least one that leads  to $P_\pm({\bm k})$ that are globally smooth over the entire Brillouin torus?

It appears to us that, for the model of Eq.~\ref{model}, the only way to construct such splitting is to use the $\hat{s}_z$ operator ($\hat{s}_z |{\bm i},\alpha,\sigma\rangle = \frac{1}{2}\sigma |{\bm i},\alpha, \sigma \rangle$). The key idea is to use the spectral properties of the operator $P\hat{s}_z P$. Indeed, after the Bloch fibration, $UP\hat{s}_z PU^{-1}$ becomes $\bigoplus_{{\bm k}\in {\cal T}}P({\bm k})\hat{s}_z P({\bm k})$ and we can diagonalize each of the operators  $P({\bm k})\hat{s}_z P({\bm k})$. If the Rashba term doesn't exceed a threshold value, the spectrum of $P({\bm k})\hat{s}_z P({\bm k})$ consists of two isolated eigenvalues, positioned symmetrically and away from the origin, for all the ${\bm k}$'s of the Brillouin torus. The spectral projectors $P_\pm({\bm k})$ onto the positive/negative eigenvalues are smooth of ${\bm k}$ and can be used to achieve the decomposition $P({\bm k})=P_-({\bm k})\oplus P_+({\bm k})$. At this point we can can define the Chern numbers $C_\pm$ for the families of projectors $P_\pm({\bm k})$ and define the spin-Chern number as $C_s$=$\frac{1}{2}(C_+$$-$$C_-)$. For the model of Eq.~\ref{model}, $C_s$={1} (assuming $\lambda_{SO}$ positive).

The smoothness of $P_\pm({\bm k})$ follows from a direct computation for the particular model of Eq.~\ref{model}, but a more general, model independent, proof was achieved in Ref.~\cite{Prodan:2009oh} based on the fact that global smoothness of $P_\pm({\bm k})$ implies exponential decay of the real space kernel $P_\pm({\bm i},{\bm j})$ with the separation $|{\bm i}-{\bm j}|$, and viceversa, exponential decay implies global smoothness. The exponential decay of $P_\pm({\bm i},{\bm j})$ can be established using traditional methods \cite{Prodan:2006of}.

Besides allowing for an explicit check of the global smoothness, the above construction also allows one to write the spin-Chern number without using the Brillouin torus, something that is absolutely necessary when dealing with disorder. Indeed, the Chern numbers $C_\pm$ can be computed in real space using the following formula, which is completely equivalent to the ${\bm k}$ space formula: 
\begin{equation}\label{realChern}
C_\pm=2\pi i \ \mbox{tr}\{P_\pm\left [ [\hat{x}_1,P_\pm],[\hat{x}_2,P_\pm] \right ]\},
\end{equation}
where ``tr" refers to the trace over the four quantum states of the first unit cell, $[,]$ denotes the usual commutator and $\hat{{\bm x}}$ is the position operator: $\hat{{\bm x}}|{\bm i},\alpha,\sigma\rangle = {\bm i}|{\bm i},\alpha,\sigma \rangle$. $P_\pm$ are the spectral projectors onto the positive and negative parts of the spectrum of $P\hat{\sigma}_zP$, respectively. This formula can be used to define the Chern numbers for the disordered case. Indeed, if we add a random potential $V_\omega$ to $H_0$, one can compute the projector $P_\omega$ onto the occupied states of $H_\omega$=$H_0$+$V_\omega$ and then investigate the eigenvalue spectrum of $P_\omega\hat{\sigma}_zP_\omega$. For $\lambda_R$ not exceeding a threshold value, one will find that the spectrum segregates into two isolated islands, one above and one below zero. Therefore, one can define the spectral projectors $P_\omega^\pm$ onto the upper/lower islands of spectrum of $P_\omega\hat{\sigma}_zP_\omega$ and compute $C_\pm$ using Eq.~\ref{realChern}. A fundamental result in noncommutative geometry \cite{BELLISSARD:1994xj} states the following:\smallskip

\noindent {\bf Proposition 1.} Let $\Pi_\omega$ be a family of projectors depending on the random variable $\omega$. If the matrix element $\langle {\bm i}|\Pi_\omega|{\bm j}\rangle$ decays sufficiently fast with the separation $|{\bm i}-{\bm j}|$ (see the following discussion) and if $\Pi_\omega$'s satisfy $u_{\bm n}\Pi_\omega u_{\bm n}^{-1}=\Pi_{ t_{\bm n}\omega}$, with $u_{\bm n}$ being the lattice translation by an arbitrary ${\bm n}$ and $t_{\bm n}$ a flow in the space of the random variable, then:
\begin{equation}
C=2\pi i \left \langle \mbox{tr}\{\Pi_\omega[[\hat{x}_1,\Pi_\omega],[\hat{x}_2,\Pi_\omega]]\} \right \rangle_\omega, \ (\langle \ \rangle_\omega = \mbox{disorder average})
\end{equation}
is an integer that is invariant to smooth deformations of $\Pi_\omega$'s as long as they remain localized.\medskip

Ref.~\cite{Prodan:2009oh} showed that, if $P_\omega$ is exponentially localized and the gap of $P_\omega\hat{s}_zP_\omega$ remains open, then $P_\omega^\pm$ are exponentially localized. The gap of $P_\omega\hat{s}_zP_\omega$ was shown to remain open as long as $\lambda_R$ is smaller than a threshold value and the disorder is not very strong. Moreover, since the spin operator commutes with translations, the relation $u_{\bm n}P_\omega^\pm u_{\bm n}^{-1}=P_{ t_{\bm n}\omega}^\pm$ is automatically satisfied, therefore $P_\omega^\pm$ satisfy the conditions of Proposition 1 and consequently $C_\pm$ and $C_s$ are topologically invariant integers as long as $P_\omega$ is exponentially localized and the gap of $P_\omega\hat{s}_zP_\omega$ remains open.

\subsection{Physical consequences}

The spin-Chern number can be used for topological classification but it will be equally important to extract non-trivial physical consequences from its remarkable properties, in particular, to say something about the localization of the bulk states in the presence of disorder. For the model defined by Eq.~\ref{model}, we can establish the following fact: a non-zero spin-Chern number implies existence of energy regions where the localization length of the bulk states diverges, at least for small $\lambda_R$. In other words, there are bulk states that resist localization in the presence of disorder.\medskip 

Indeed, let us imagine a numerical experiment in which we lower the Fermi level continuously, from the mid-gap all the way to $-\infty$, and try to repeat the construction of the spin-Chern number, this time starting from $P_\omega(E_F)\sigma_zP_\omega(E_F)$ [$P_\omega(E_F)$ = the projector onto the energy spectrum below $E_F$]. When the Fermi level enters the bulk energy spectrum, the spectrum of $P_\omega(E_F)\sigma_zP_\omega(E_F)$ can rapidly spread within the entire $[-\frac{1}{2},\frac{1}{2}]$ interval, engulfing the origin. Even though, according to the general results of Ref.~\cite{BELLISSARD:1994xj}, the Chern numbers for projectors $P_\omega^\pm(E_F)$ remain well defined and invariant of $E_F$ as long as the spectrum of $P_\omega(E_F)\sigma_zP_\omega(E_F)$ remains localized near the origin, more precisely as long as:
\begin{equation}\label{cond}
\left \langle  \mbox{Tr}\{\pi_0 P_\omega^\pm(E_F) {\bf x}^2 P_\omega^\pm(E_F) \pi_0\} \right \rangle_\omega < \infty,
\end{equation}
where $\pi_0$ is the projector onto the four states of the first unit cell. This condition says that the kernel 
\begin{equation}
P_{\omega,E_F}^\pm(0,\alpha,\sigma;{\bm i},\alpha',\sigma')=\langle 0,\alpha,\sigma |P_\omega^\pm(E_F) |{\bm i},\alpha',\sigma'\rangle\end{equation}
decays fast enough so that:
\begin{equation}
\Lambda^\pm(E_F) =\left [ \sum_{\bm i} |{\bm i}|^2 \sum_{\alpha,\sigma}\sum_{\alpha',\sigma'}  |P_{\omega,E_F}^\pm(0,\alpha,\sigma;{\bm i},\alpha',\sigma') |^2 \right ]^{1/2}
\end{equation}
is finite. The quantity inside the square root is precisely the left side of Eq.~\ref{cond}. $\Lambda^\pm(E_F)$ can be viewed as (polarized-) localization lengths. 

Our real interest is in the true localization length:
\begin{equation}
\Lambda(E_F) = \left [\left \langle  \mbox{Tr}\{\pi_0 P_\omega(E_F) {\bf x}^2 P_\omega(E_F) \pi_0\} \right \rangle_\omega \right ]^{1/2}.
\end{equation}
The connection between the true and the polarized localization lengths is given by the following relation:
\begin{equation}\label{inter}
\begin{array}{l}
\Lambda(E_F)^2=\Lambda^+(E_F)^2+\Lambda^-(E_F)^2\medskip\\
+\left \langle  \mbox{Tr}\{\pi_0 P_\omega^+(E_F) {\bf x}^2 P_\omega^-(E_F) \pi_0\} \right \rangle_\omega \medskip \\
 +\left \langle  \mbox{Tr}\{\pi_0 P_\omega^-(E_F) {\bf x}^2 P_\omega^+(E_F) \pi_0\} \right \rangle_\omega .
\end{array}
\end{equation}
But the last two lines are null when $\lambda_R$=0, and they remain small relative to the terms appearing in the first line, for small $\lambda_R$. 

Now as we move the Fermi level to $-\infty$, the Cern numbers $C_\pm$ become zero, therefore they must jump at some point from $\pm 1$ to zero. This can happen only if $\Lambda^\pm(E_F)$ diverge as $E_F$ crosses  certain energy regions. According to Eq.~\ref{inter},  this implies that the true localization length  $\Lambda(E_F)$ also diverges when $E_F$ crosses these energy regions, at least for small $\lambda_R$.

\subsection{Discussion}

 We would like first to relate our findings to the numerical studies of Refs.~\cite{Sheng:2006na,Onoda:2007xo}. In the following, if not otherwise specified, the Fermi level is considered in the middle of the bulk gap. The first observation is that the original construction in Ref.~\cite{Prodan:2009oh} considered same type of disorder as in these references, a white noise with amplitude between $[-W/2,W/2]$ (to keep the notation of  Refs.~\cite{Sheng:2006na,Onoda:2007xo}]). For the model of Eq.~1, by direct computation in the clean limit, we have verified that the gap of $P_\omega\hat{s}_zP_\omega$ remains open when $\lambda_R$ is increased, all the way until the gap of $H_0$ closes. This tells us that $C_s$ is a robust topological invariant, for the entire domain of QSH phase, in total agreement with the numerical observations of \cite{Sheng:2006na}. In Ref.~\cite{Prodan:2009oh}, the amplitude $W$ was assumed small so that a gap remains open in the spectrum of $H_\omega$, so this work established the robustness of $C_s$ only in the presence of relatively small disorder. The study by Sheng and her collaborators went beyond the small disorder case and established that $C_s$ remains well defined even if the insulating gap is filled with localized spectrum.
 
Switching to the second numerical study \cite{Onoda:2007xo}, we would like to specify that  the discussion of Ref.~\cite{Prodan:2009oh} was limited to the range $W\leq 3$ of Ref.~\cite{Onoda:2007xo}. In this case, the exponential localization of $P_\omega$ and of $P_\omega^\pm$ can be explicitly proven; therefore, according to our arguments, one should observe bulk states that resist localization, in line with the numerical findings of Ref.~\cite{Onoda:2007xo}. Beyond $W=3$, the spin-Chern number remains a topological invariant only if the Fermi level can be placed in a region of localized states, as it was already discussed above. Referring to Fig.~1 of  Ref.~\cite{Onoda:2007xo}, this is the case for panels (a-1), (b-1) and (c-1), which correspond to $W=5$ and different values of $\lambda_R$, but not for panels (a-2), (b-2) and (c-2) which correspond to $W$=7. $C_s$ can be defined again for panels (a-3), (b-3) and (c-3), correspoinding to $W$=8, but this time it takes a trivial value. It is interesting to notice that the evolution of the de-localized spectral regions with the increase of $W$ seen in these numerical experiments can be actually predicted using the following arguments. As long as the Fermi level can be placed in a region of localized spectrum, the spin-Chern number remain non-trivial, therefore we should observe de-localized states below and above $E_F$. It is known \cite{Aizenman:1993ax}, however, that for very large values of $W$, all the states localize, therefore at some point there should be no alternative, but to place $E_F$ in a de-localized region of the spectrum. But $E_F$ can be always placed in a region of localized spectrum unless the de-localized spectral regions below and above $E_F$ move towards each other and at some point "scissor" the Fermi level. This is precisely what was observed in Ref.~\cite{Onoda:2007xo}. 

Secondly, we would like to discuss some predictions. The Z$_2$ topological invariant can be computed as $C_s$(mod2). According to our arguments, there should be bulk states that resist localization not only for odd but also for even spin-Chern numbers, that is, for trivial Z$_2$ invariants. This, for example, must be the case for the model of Eq.~\ref{model} with spin $\frac{3}{2}$, which has $|C_s|$=2. Therefore, it will be very interesting to see the calculations of Ref.~\cite{Onoda:2007xo} repeated for this system. 

Another observation is that the spin-Chern number can be defined even in the absence of time-reversal symmetry. Ref.~\cite{Onoda:2007xo} presented a calculation that included a Zeeman term in the $x$ direction, of strength $h_x$. The results for relatively large magnetic field indicate that the bulk states are entirely localized (although the graph corresponding to the weaker disorder is not very conclusive in our opinion). Unfortunately, Ref.~\cite{Onoda:2007xo} does not discuss what happens at weaker magnetic fields, i.e. when the time reversal symmetry is only weakly broken. In this regime, the last two lines of Eq.~\ref{inter}  remain small and a divergence of $\Lambda^\pm(E_F)$ implies a divergence of $\Lambda(E_F)$. In other words, we predict that, for small values of $h_x$, there are bulk states that resist localization in the presence of disorder.

Our final remark for the section is that the present construction of the spin-Chern number can be repeated for other operators, let us call them $\hat{w}$, instead of $\hat{s}_z$.  If the eigenvalue spectrum of $P\hat{w}P$ consists of two or more isolated islands, one can compute the Chern numbers of the spectral projectors for each island and generate $\hat{w}$-Chern numbers. A non-zero $\hat{w}$-Chern number is a signal of topological properties, leading to a possible $\hat{w}$-Hall insulator. We hope that this observation will lead to more efficient search and discovery paths for new classes of topological insulators. 

\section{Tools for the edge states}

In 2004, the following result appeared \cite{Kellendonk:2004et}. Consider the Landau Hamiltonian $H_\omega$ describing free electrons in the semiplane $z$=0 and $x$$>$0 (therefore with an edge along $y$ axis), moving in a periodic potential, a uniform magnetic field along the z axis (no rational flux required) and a random potential $V_\omega$. Then the following is true:
\begin{equation}\label{Kell}
\left \langle \mbox{tr}_0\left \{\rho(H_\omega)\frac{d\hat{y}}{dt}\right \}\right \rangle_\omega = \frac{1}{2\pi} \mbox{Ind} \{ \chi_+ e^{-2\pi i F(H_\omega)}\chi_+ \}
\end{equation}
where $\rho(\epsilon)$ is any arbitrary distribution with support in a bulk energy gap of $H_\omega$ and such that $\int \rho(\epsilon)d\epsilon$=1, the function $F$ is defined by $F(\epsilon)$=$\int^{\infty}_\epsilon \rho(s) ds$ and $\chi_+(x,y)$ is the characteristic function for the semi-plane $y$$>$0. The trace tr$_0$ is taken over the states with support in the plane $y$=0. 

This is a remarkable result since it contains the first explicit proof that the edge states of the Landau Hamiltonian resist localization in the presence of disorder. Indeed, let us read again Eq.~\ref{Kell}. The left hand side is the total charge current of the edge states inside a bulk energy gap of $H_\omega$, weighted with the distribution $\rho$. This current will be zero if the edge states localize. Therefore, if one can show that the right hand side is non-zero, he has an explicit proof that the edge states resist localization. But the right hand side is the index of an operator (to be discussed shortly), which can be computed explicitly and for the Landau Hamiltonian it turns out to be a non-zero integer number. One can see then how Eq.~\ref{Kell} becomes a very effective tool for the study of the edge physics.

The above result is relevant for Landau Hamiltonians and for Chern insulators \cite{Prodan:2009lo}, but unfortunately is not relevant for time-reversal invariant topological insulators because their edge charge current is zero. Therefore, we need to look at the currents of other observables (spin, etc.). This was the main motivation behind the work of Ref.~\cite{Prodan:2009od}. Using non-commutative calculus, the author established a general result stating when and why is the current of a general observable $X$ quantized. The result can be summarized as follow. Assume:\medskip

\noindent a) Existence of a family of self-adjoint Hamiltonians $H_\omega$ on $\cal{H}$, with $\omega \in \Omega$. 

\noindent b) Existence of a 1-parameter unitary group $u_a$ such that $u_a X u_a^{-1}=X+a$.

\noindent c) Existence of an ergodic flow $t_a$ on $\Omega$ such that $u_a H_\omega u_a^{-1} = H_{t_a \omega}$ for all $\omega \in \Omega$.\medskip

\noindent {\bf Proposition 2.} If $\pi_+$ denotes the projector onto the positive eigenvalue spectrum of $X$, $\rho(\epsilon)$ is a statistical distribution of the quantum states such that $\int  \rho(\epsilon)d \epsilon$=1 and  $F(\epsilon)$=$\int_\epsilon^{\infty} \rho(\epsilon) d \epsilon$, then: [below, $ X_\omega(t)=e^{itH_\omega}X e^{-itH_\omega}$]

\begin{equation}\label{TopQ}
\left \langle \mbox{tr}_0 \left\{\rho(H_\omega) \frac{\mbox{d}X_\omega(t)}{\mbox{d}t}\right\} \right \rangle = \frac{1}{2\pi}   \mbox{Index} \left \{\pi_+ e^{-2\pi i F(H_\omega)} \pi_+ \right \},
\end{equation} 
provided the kernel of $e^{-2\pi i F(H_\omega)}$ has certain localization properties.\medskip 

\noindent The index of an operator is equal to the difference between the number of zero modes of that operator and the number of zero modes of its conjugate. It is an integer number, invariant to continuous deformations of the operator. The index on the right hand side of Eq.~\ref{TopQ} is now called the edge index of observable $X$, and can be explicitly computed by deforming the Hamiltonian into simple, exactly solvable models. During such deformations, one has to make sure that the technical conditions are not violated, that is, the localization of $e^{-2\pi i F(H_\omega)}$.

Eq~\ref{TopQ} can be a very useful tool for the analysis of the edge spectrum. Note that the left side of Eq.~\ref{TopQ} is zero if the spectrum of $H_\omega$ is completely localized. Therefore, a non-zero value of the right side of Eq.~\ref{TopQ} assures that $H_\omega$ has absolutely continuum spectrum, even in the presence of disorder.  It becomes clear then that Eq.~\ref{TopQ} establishes a very general method to demonstrate the robustness of the edge spectrum against disorder, which amounts to a search of an observable $X$ for which the index of the right hand side of Eq.~\ref{TopQ} is non-zero.

For Quantum Spin-Hall insulators, this observable was defined in Ref.~\cite{Prodan:2009mi}.  The construction of the observable $X$ was  based on the fact that for spin $\frac{1}{2}$ particles and time-reversal invariant Hamiltonians, the Hilbert space can be split into two sectors ${\cal H}$=${\cal H}_-$$\oplus$${\cal H}_+$ such that each of the sectors are left invariant by the Hamiltonian and $\tau{\cal H}_\pm$=${\cal H}_\mp$ ($\tau$ = time-reversal operation). The observable $X$ is simply $X$=$P_+yP_+$$-$$P_-yP_-$, where $P_\pm$ are the projectors onto the sectors ${\cal H}_\pm$ and $y$ is the coordinate along the edge. For the model of Eq.~\ref{model} with an edge and random potential, the edge index of this observable is equal to spin-Chern number. In other words, using the above tools, we now have an explicit proof that the edge states of Quantum Spin-Hall insulators resist localization when disorder is present.

The original and now most popular argument \cite{Kane:2005np} for the robustness of the edge states is based on the observation that the scattering between the right and left moving edge bands cannot occur in time-reversal invariant insulators. In author's opinion, this argument is not enough. The reason is because, although we are concerned with the quasi-one dimensional edge states, we are still dealing with a 2D systems. In such systems, a band can scatter into an infinite number of analytically continued bands as detailed in Refs.~\cite{Prodan:2006pt} and \cite{Prodan:2009hi}. In other words, if one takes an energy $\epsilon$ in the bulk insulating gap and $\epsilon_{n,k}$ denotes the band energies of the model with an edge, the equation $\epsilon$=$\epsilon_{n,k}$ has an infinite number of solutions, thought many of them will occur at complex $k$. Nevertheless, the incoming edge state $\psi_{\mbox{\tiny{in}}}$ will scatter into these complex $k$ states, which are the analytic continuation of the bulk states, and time-reversal invariance cannot prevent this. For these reasons, it appears to the author that the edge bands can backscatter through the bulk.

\section{Outlook and concluding remarks}

We hope that the present paper makes more clear the motivation behind the somewhat abstract tools developed with the aid of non-commutative calculus and geometry. They allow one to define robust topological invariants directly in the thermodynamic limit and in the presence of disorder. Their robustness can have direct physical implications and the non-commutative methods can lead to precise statements about the localization of the bulk and the edge states. The list of known topological insulators is expected to increase in the following years and the generalization of the spin-Chern number to the $\hat{w}$-Chern number could aid the search and discovery of new classes of topological insulators. The edges of these new topological insulators could cary currents of new observables, therefore the generality of the tools developed for the edge physics may become more useful in the future.

One of the interesting and important things left unsolved is how to define the Z$_2$ bulk topological number without involving the Brillouin torus and how to construct its non-commutative theory? What are the physical implications of its robustness against disorder for the bulk states? In the author's view, these are the most pressing questions in the field of topological insulators.

As one can see, the discussion of the paper did not touch the three dimensional case at all. Unfortunately, the applications of non-commutative geometry to 3D systems are very scarce. It is author's hope that the discovery of the 3D topological insulators and the recent and exciting theoretical work on these systems will renew the interest of non-commutative geometers and new applications will come soon.\medskip

\bibliographystyle{iopart-num} 
\providecommand{\newblock}{}

\end{document}